\documentclass[fleqn,usenatbib]{mnras}
\usepackage{newtxtext,newtxmath}
\usepackage[T1]{fontenc}
\DeclareRobustCommand{\VAN}[3]{#2}
\let\VANthebibliography\thebibliography
\def\thebibliography{\DeclareRobustCommand{\VAN}[3]{##3}\VANthebibliography}
\usepackage{graphicx}	
\usepackage{amsmath}	
\usepackage{multicol}

\title[Evolution of Amati parameters]{Investigating the Evolution of Amati Parameters with Redshift \vspace{-2ex}}
\author[ Meghendra Singh et al.]{\large Meghendra Singh,$^{1}$ Darshan Singh,$^{2}$ Kanhaiya Lal Pandey,$^{3}$ Dinkar Verma,$^{2}$ and Shashikant Gupta$^{2}$
\thanks{E-mail: shashikantgupta.astro@gmail.com(SG)}
\\
$^{1}$Delhi Metro Rail Corporation Limited, New Delhi, 110001, India\\
$^{2}$GD Goenka University, Gurugram, 122103, India\\
$^{3}$Department of Physics, School of Advanced Sciences, Vellore Institute of Technology, Vellore, 632014, India
}
\date{Accepted XXX. Received YYY; in original form ZZZ}
\pubyear{2022}
\begin{document}
\label{firstpage}
\maketitle

\begin{abstract}
Gamma Ray Bursts (GRB) are among the brightest objects in the Universe and hence can be observed up to a very high redshift. Properly calibrated empirical correlations between intensity and spectral correlations of GRBs can be used to estimate the cosmological parameters. However, the possibility of the evolution of GRBs with the redshift is a long-standing puzzle. In this work, we used 162 long-duration GRBs to determine whether GRBs below and above a certain redshift have different properties. The GRBs are split into two groups, and we fit the Amati relation for each group separately. Our findings demonstrate that estimations of the Amati parameters for the two groups are substantially dissimilar. We perform simulations to investigate whether the selection effects could cause the difference. Our analysis shows that the differences may be intrinsic, and the selection effects are not their true origin.
\end{abstract}


\section{Introduction}
\label{sec:intro}
Gamma-ray bursts (GRBs) discovered in 1967 \cite{Klebesadel1973} are the most powerful explosions in the Universe. The Compton Gamma Ray Observatory's Burst and Transient Source Experiment (BATSE) confirmed the isotropic distribution of GRBs \citep{meegam92,briggs1996}, indicating that they are extra-galactic. GRBs are short bursts of gamma-ray flare which last from milliseconds to a few minutes. GRBs produce as much energy in a few seconds as the Sun does in its lifespan. Observations suggest two types of GRBs, short and long, based on $T_{90}$, the duration in which $90$ percent of the burst energy is emitted. Short GRBs with $T_{90} < 2$ s are thought to be resulted from compact star mergers. On the other hand, Long GRBs are thought to arise from a massive star's core collapse, with a length of $T_{90} > 2$ s. \citep{Woosley1993,Paczynski1998, MacFadyen1999}.

The ability to identify GRBs up to extremely high redshifts has long piqued the interest of cosmologists. Long gamma-ray bursts (GRBs) provide a unique opportunity to investigate the properties of galaxies at high redshift, including galaxy evolution, star formation rate, and the intergalactic medium (IGM) and interstellar medium (ISM) \citep{Woosley2006,Prochaska2007,Fynbo+2009,Tanvir2009,Salvaterra2009,Cucchiara+2011}. In the last two decades, various relations between different observational parameters have been discovered, such as the Amati relation, and Ghirlanda relation \citep{ghi04,lam05,fir06,blo03,fri05,Schaefer2007}. These relations have been used to calibrate secondary distance indicators.

The Amati relation is one of the most important relations in the context of GRBs. It's a correlation in the $\nu F_{\nu}$ spectrum between a GRB's isotropic-equivalent energy ($E_\mathrm{iso}$) and its intrinsic peak energy ($E_\mathrm{p,i}$). It was discovered in 2002 \citet{ama02}, and subsequent research has confirmed it \citep{ama06,ama08,ama09}. If the redshift, $z$, of the GRB is known, the $E_\mathrm{p,i}$ can be estimated from the observed peak energy, $E_\mathrm{p ,obs}$, using the relation:
\begin{equation}
E_\mathrm{p,i} = (1+z) E_\mathrm{p ,obs}\ . 
\label{eq:epi}
\end{equation}
The Amati relation is then given by :
\begin{equation}
E_\mathrm{p,i} = K \left( \frac{E_\mathrm{iso}}{10^{52} \rm{erg}} \right)^{m}\, , \label{eq:amati1}
\end{equation}
where $K$, and $m$ are constants, and {$E_\mathrm{p,i}$} is expressed in keV. In Amati's original study, $m$ and $K$ were determined to be $m\approx 0.5$ and $K\approx 95$, respectively \cite{ama02}. Later research using bigger samples of long GRBs supports the aforementioned range of $K$, and $m$ estimations. Alternatively, the Amati relation can be expressed in the logarithmic form as:
\begin{eqnarray}
     \log E_{\mathrm{iso}} = a + b \log E_{\mathrm{p,i}} \;. 
     \label{eq:amati2}
\end{eqnarray}
Eq~\ref{eq:amati2} has the advantage of being linear. By taking logarithms on both sides, Eq~\ref{eq:amati1} can be written as: 
\begin{equation}
\log E_\mathrm{iso} = \frac{1}{m} \log E_\mathrm{p,i} - \frac{1}{m} \log K \ . \label{eq:amati3}
\end{equation}
Comparing Eq~\ref{eq:amati2} and Eq~\ref{eq:amati3}, one can express $a$ and $b$ in terms of $K$ and $m$  as
$a = - \frac{1}{m} \log K \,$ and $b = \frac{1}{m}$.

There are some issues with the Amati relation as well \citet{Andrew2012}. One such issue is the redshift degeneracy in the $E_\mathrm{iso}-E_\mathrm{ peak}$ relation \citet{Li06}. Another problem is that approximately $48$ percent of the GRBs in the BATSE data violate the Amati relation \cite{nakar2005}. It's also worth noting that hybrid samples with a wide range of redshifts are frequently used while fitting GRB data for correlations. As a result, the evolution of GRBs with redshift and the selection effects are often ignored.

It is counter-intuitive to neglect the evolution of GRBs in the large redshift range [$0.1 \le z \le 9$]. Long-duration GRBs are more likely to occur in galaxies with low metallicity \cite{fyn03,hjo03,lef03,sol05,fru06}. The isotropic energy of long GRBs is anti-correlated with the host galaxy's metallicity \cite{sta06}. Metallicities are widely known to vary with cosmic redshift \cite{kew05,sav05}. According to \cite{lan06}, GRB evolution is expected as redshift increases. It makes sense to investigate the redshift dependence of the relations found in the GRB samples. Using 48 long GRBs \cite{Li2007} investigated the redshift dependency of the Amati parameters $a$ and $b$ with the help of $48$ long GRBs. These GRBs were divided into four redshift bins, and a linear variation in both $a$ and $b$ was observed. The authors claim that the variation could not be due to the selection effects.
\citet{Dainotti2013} studied the origin of correlation between the X-ray luminosity (Lx) and the rest frame time (Ta). \cite{Dainotti2015} also investigated whether the redshift-dependent ratio of GRB rate to star formation evolves with redshift. Their results indicate a modest evolution of this ratio in the range $0.99 < z < 9.4$. 
The luminosity of host galaxies in the NIR range is significantly high at redshifts below 1.5. Dust-obscured GRBs are mostly found in massive galaxies but rarely in low-mass galaxies. This indicates that massive galaxies are mostly dusty, while low-mass galaxies contain little dust in their ISM. It is well known \cite{Perley2016} that the ratio of GRB rate to star formation is almost constant in metal-poor galaxies while it is suppressed in metal-rich galaxies. This is linked to the GRB rate being low at z<1.5. \cite{tan2015} studied the evolution of luminosity function and the redshift selection effect of long GRBs. Their findings suggest that redshift detection efficiency, particularly in the range $1 < z < 2.5$, decreases with redshift giving rise to the redshift desert effect. It is to be noted that the redshift desert is found in the case of low luminosity GRBs and the brighter GRBs have no contribution to it. 
In the present paper, we investigate the evolution of the Amati parameters $a$ and $b$ with the redshift. We also investigate whether the selection biases could be responsible for the possible redshift dependence if any. The data set used in our analysis contains three times more GRBs than \cite{Li2007}. Various cosmological observations indicate quenching in galaxy clusters and a variation in star formation rate  around a redshift of $\sim 1.5$ \cite{Nantais2016,Krumholz2012,Zhiyuan2018,Rychard2020}. In light of the facts mentioned above, we plan to investigate the evolution of the Amati parameters accordingly.\\
The rest of the paper is organized as follows: We outline the GRB data set and the analysis methods in Section \S~\ref{sec:data-method}. We offer a summary of our results and a brief explanation of the selection effect in Section \S~\ref{sec:results} and present  conclusions in Section \S~\ref{sec:conclusion}.
\begin{figure}
\centering
\includegraphics[width=1.0\linewidth ]{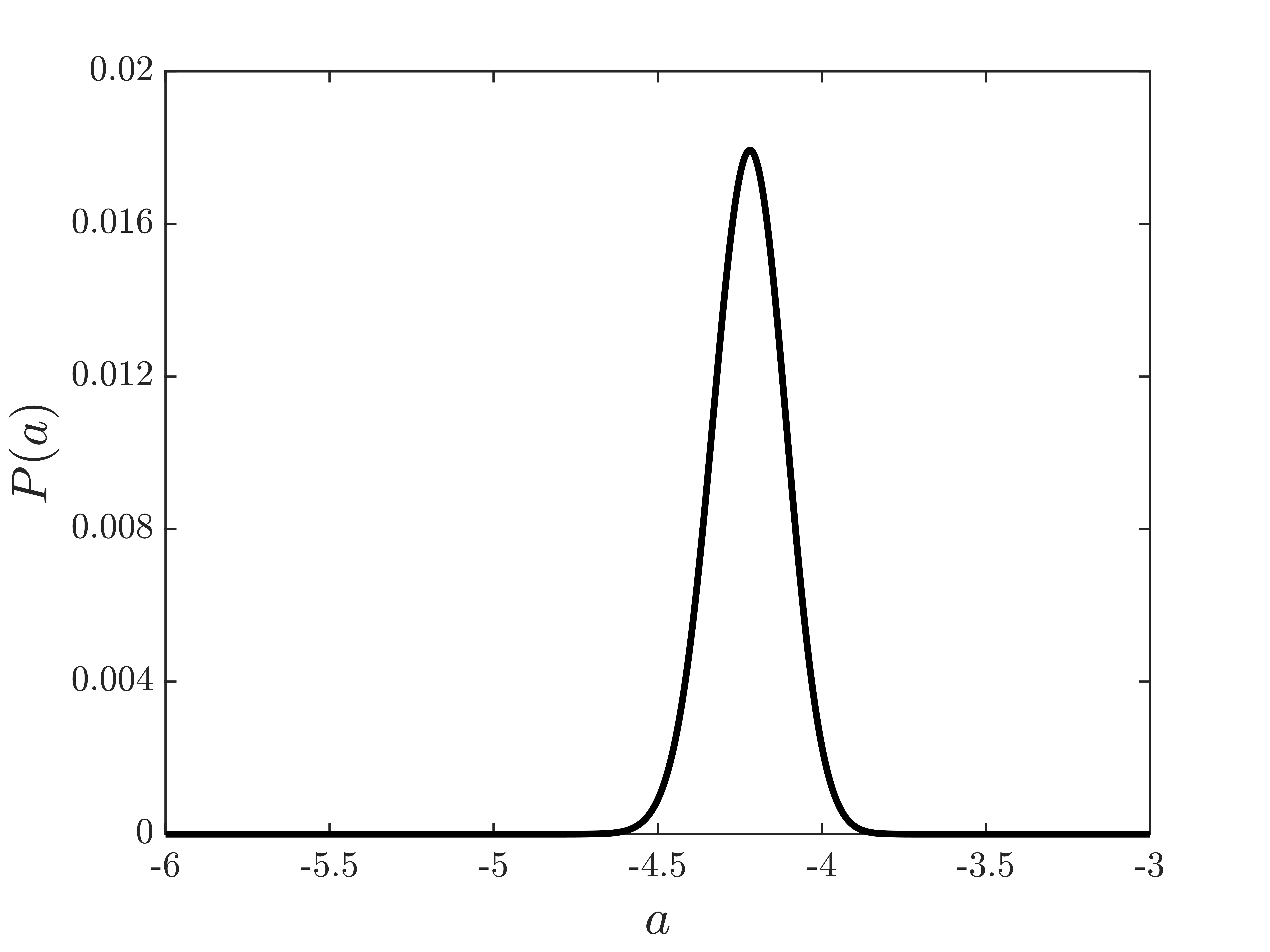}
\caption{Posterior probability of the Amati parameter, $a$ for 162 GRBs.}
\label{fig:a-full-post}
\end{figure}
\begin{figure}
\centering
\includegraphics[width=1.0\linewidth]{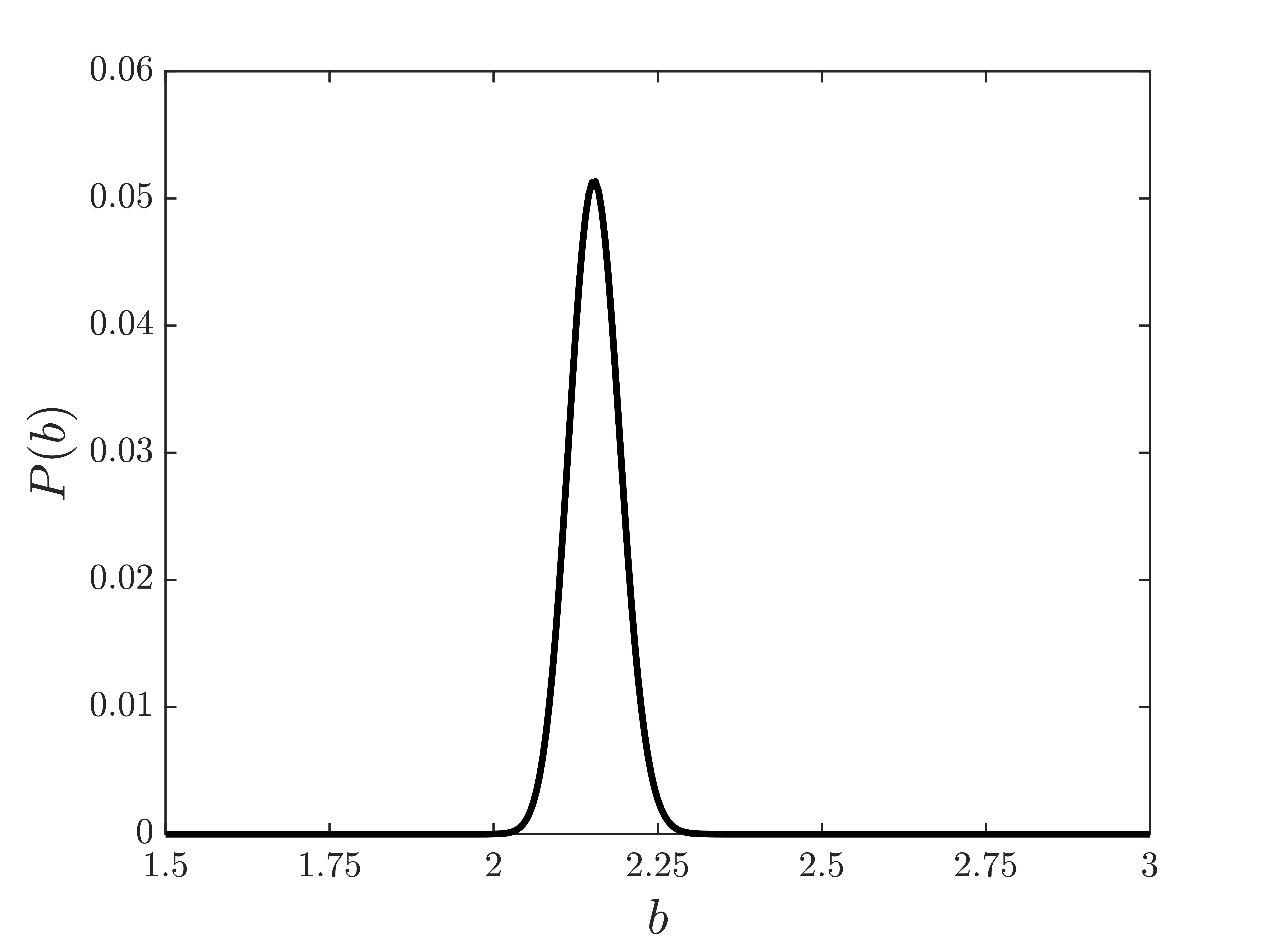}
\caption{Posterior probability of the Amati parameter, $b$ for 162 GRBs.}
\label{fig:b-full-post} 
\end{figure}
\begin{figure}
\centering
\includegraphics[width=1.0\linewidth]{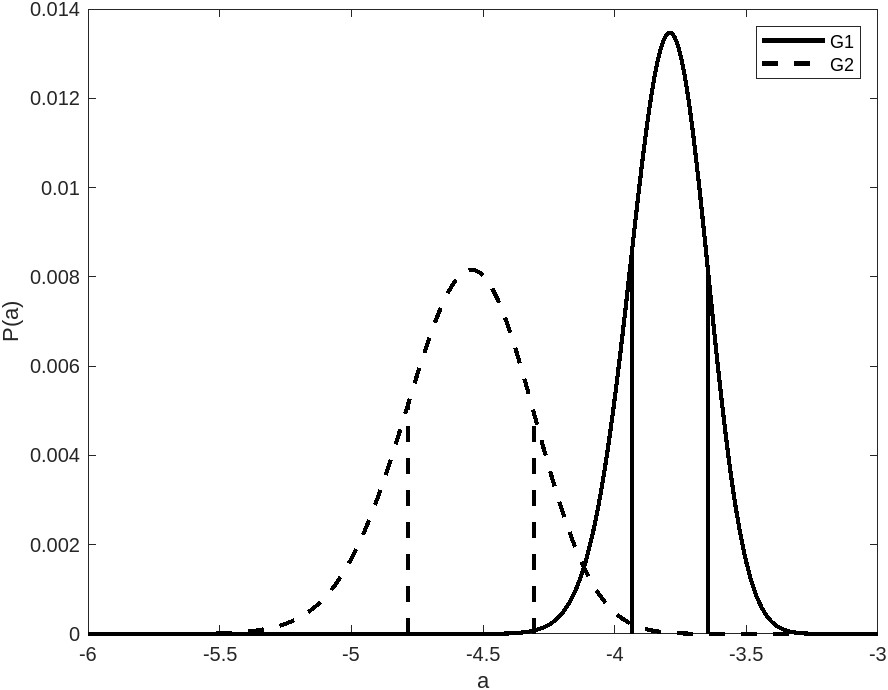}
\caption{Posterior probability curve for the Amati parameter, $a$ for groups $G_1$ and $G_2$. The values do not agree at $1.9 \sigma$.}
\label{fig:a-grup-post} 
\end{figure}
\begin{figure}
\centering
\includegraphics[width=1.0\linewidth]{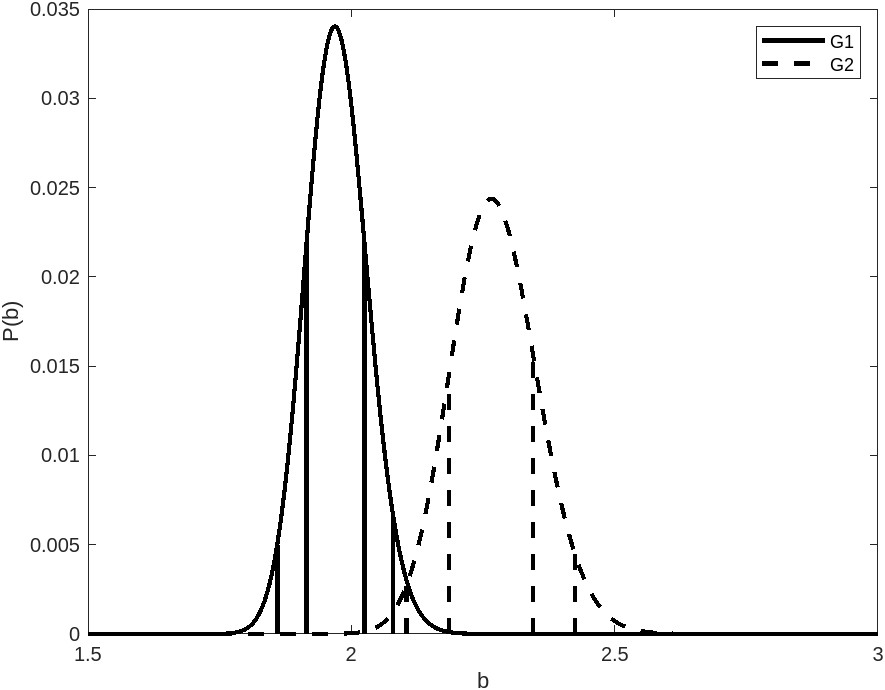}  
\caption{Posterior probability curve for the Amati parameter, $b$ for groups $G_1$ and $G_2$. The values do not agree at $2 \sigma$ level.}
\label{fig:b-grup-post}
\end{figure}
\begin{figure}
\centering
\includegraphics[width=1.0\linewidth]{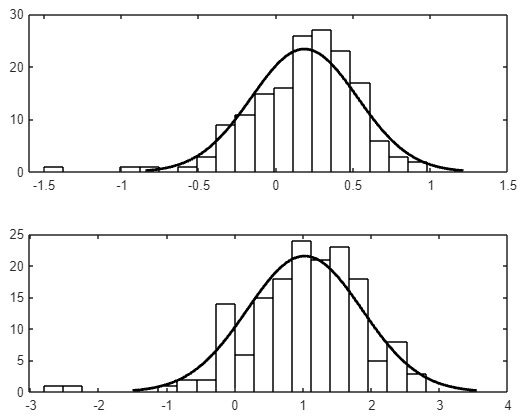}
\caption{Distribution of redshift, $z$ and isotropic-equivalent energy, $E_{iso}$ of the observed long GRBs. The log$z$ values in the data fit a log-normal distribution with a mean $0.18$ and standard deviation $0.34$ (upper panel). log$E_{iso}$ values also fit a log-normal distribution with mean $1.01$ and standard deviation $0.84$ (lower panel).}
\label{fig:lognormal}
\end{figure}
\begin{figure}
\centering
\includegraphics[width=1.0\linewidth]{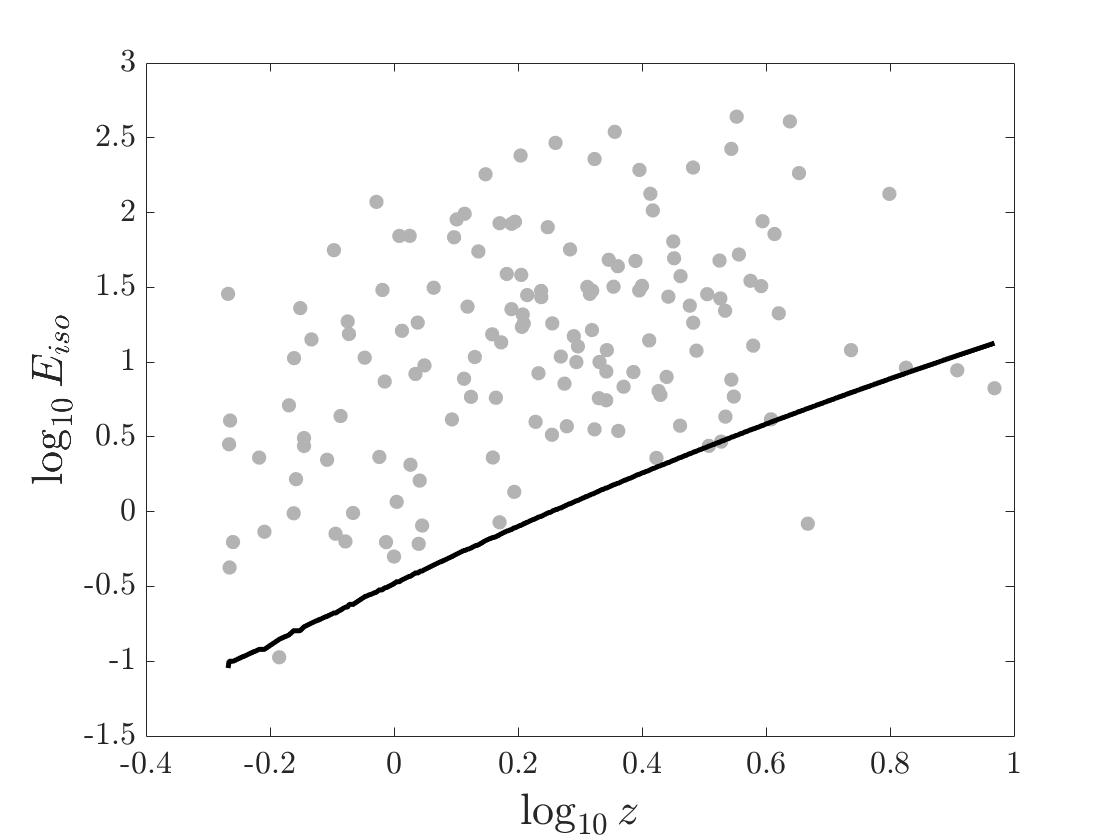} 
\caption{Variation of isotropic-equivalent energy ($E_{iso}$) with redshift ($z$) of $162$ GRBs. The solid line has been plotted using Eq.~\ref{eq:limit} with a bolometric fluence limit of $F_{bol,lim} = 10^{-9} \rm{ergs/cm^2}$. The solid line shifts upward when $F_{bol,lim}$ is increased.}
\label{fig:fbol}
\end{figure}
\begin{figure}
\centering
\includegraphics[width=1.0\linewidth]{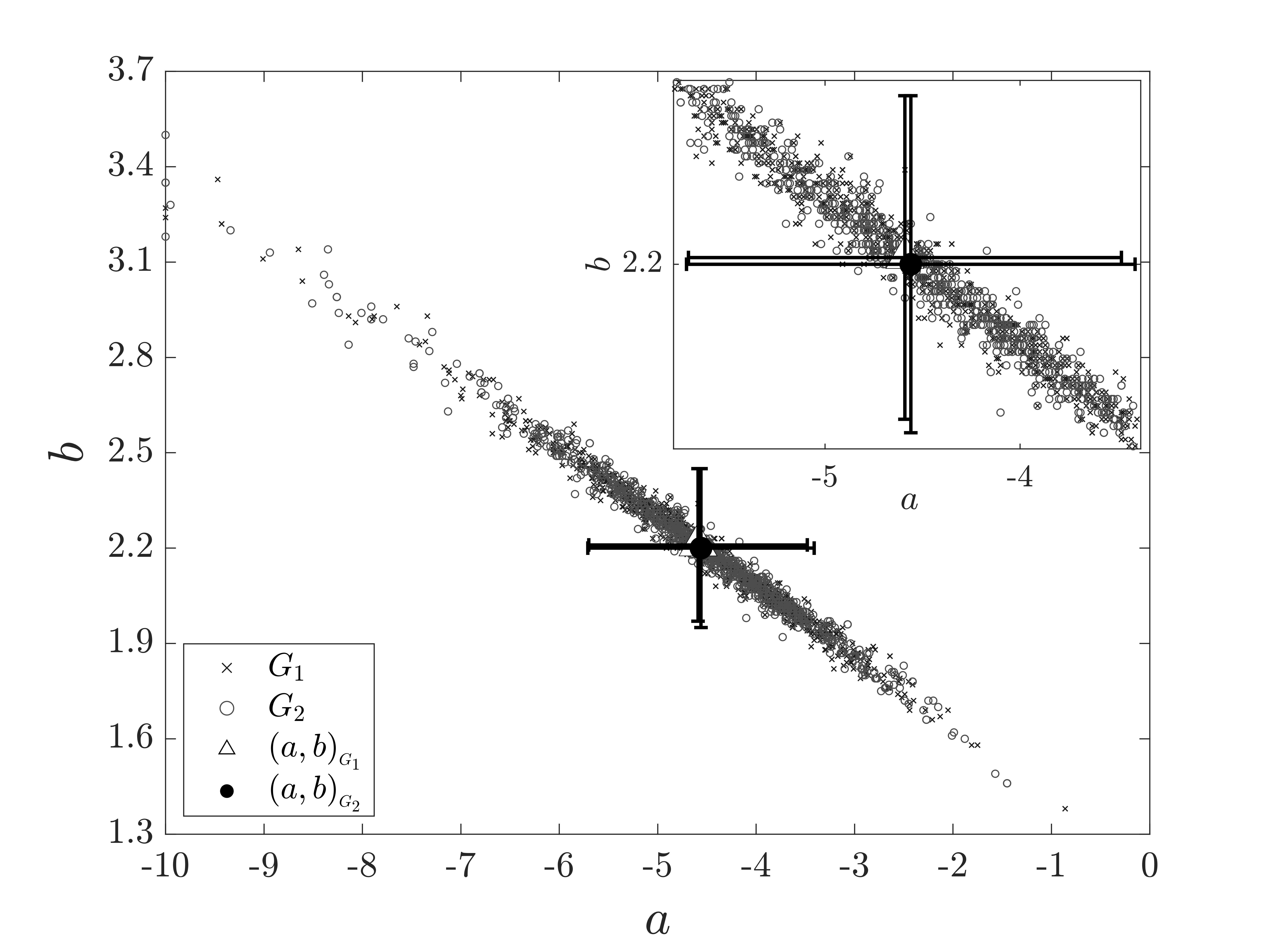}
\caption{Best-fit values of the Amati parameters, $a$ and $b$ for simulated data sets. Each data set consists of $1000$ values of $z$ and $E_{iso}$ which are further divided into low and high-$z$ groups ($G_1$ and $G_2$). A lower limit on $E_{iso}$ has been imposed using Eq.~\ref{eq:limit}. Symbol '.' represents simulated data sets for low-$z$ while 'x' represents the high-$z$ simulated data sets. Empty and dark circles show the average best-fit values of $a$ and $b$ from the low-$z$ and high-$z$ groups of simulated data.}
\label{fig:withselectioneffect}
\end{figure}
\begin{figure}
\centering
\includegraphics[width=1.0\linewidth]{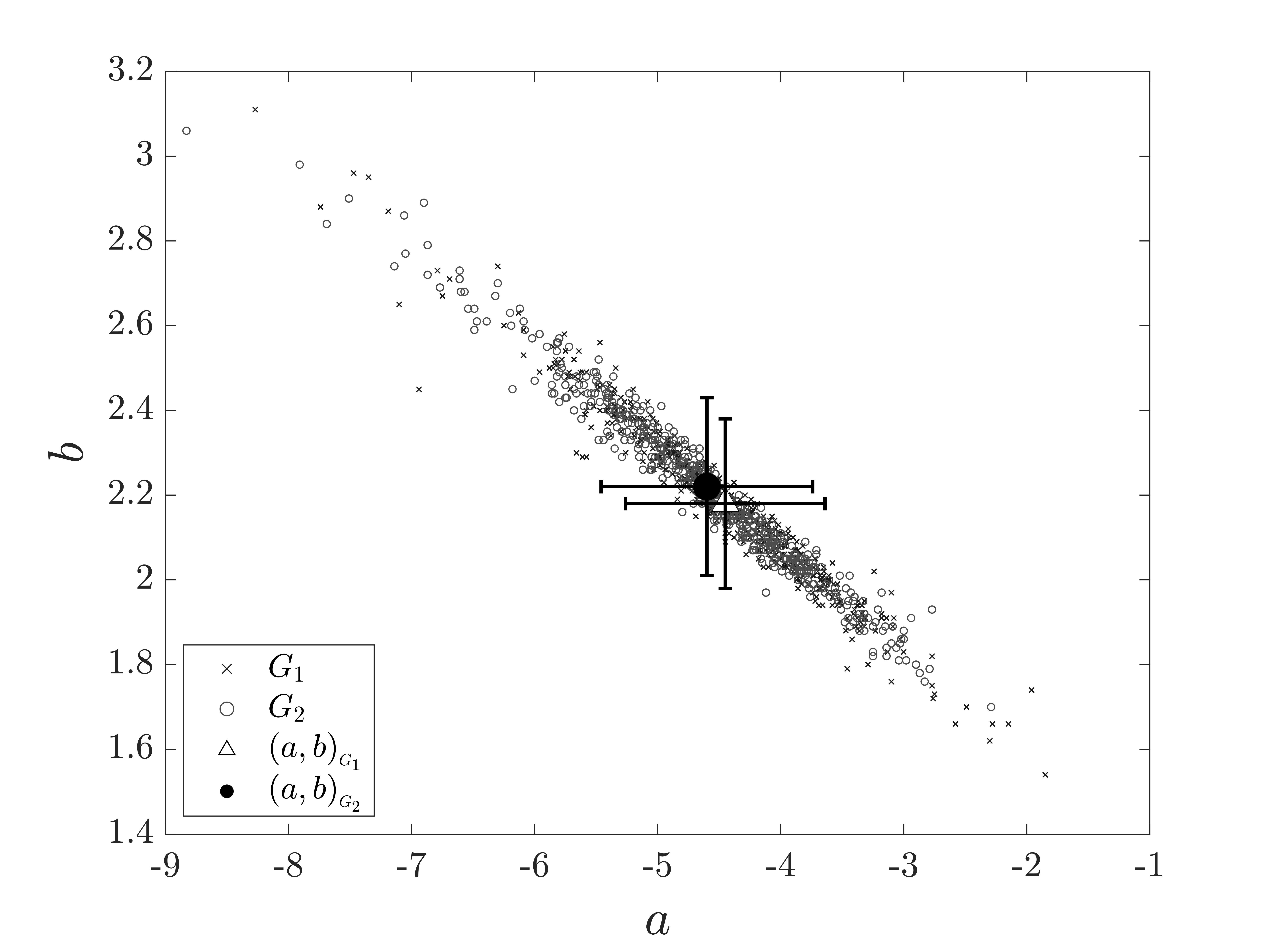}  
\caption{Best-fit values of the Amati parameters, $a$ and $b$ for simulated data sets. No lower limit has been imposed on $E_{iso}$. Symbol '.' represents simulated data sets for low-$z$ while 'x' represents the high-$z$ simulated data sets. Empty and dark circles show the average best-fit values of $a$ and $b$ from the low-$z$ and high-$z$ groups of simulated data.}
\label{fig:withoutselectioneffect}
\end{figure}
\section{Data and Methodology}
\label{sec:data-method}
\subsection{Data}
\label{sec:data}
We require a data set that encompasses a large number of GRBs and spans a wide redshift range to study the evolution of GRBs. An updated compilation of long GRB data, which contains spectrum and intensity details as well as precise information about redshift, is available \cite{Demianski17}. This data set contains 162 long GRBs and X-ray flashes (XRFs) with an extensive redshift range ($0.03 \leq z \leq 9.3$), making it suitable for our purpose. The major contribution in the data sample comes from the joint observations of
Swift and Fermi or Konus-WIND; however, in some cases, $E_{p,i}$ has been provided directly by the Swift/BAT. 
\subsection{Methodology}
\label{sec:method}
To comprehend the differences between GRB populations below and above the redshift, $z=1.5$, we especially want to look at the development of the Amati parameters with redshift. Keeping in mind that the GRB production rate in galaxies differs below and above the redshift of $1.5$ due to metallicity differences \cite{Perley2016}, we divide the GRB data into two groups of low and high redshift for this purpose. There are 71 GRBs in the low $z$ sample, with a median redshift $0.86$ and a range $0.033<z<1.489$. The high $z$ sample consists of $91$ GRBs with a redshift range $1.52<z<9.3$ and a median $2.48$.

The isotropic-equivalent energy, $E_\mathrm{iso}$ and peak energy, $E_\mathrm{p,i}$ measurements, together with the observational errors, are known for $162$ GRBs. 
To fit for the Amati parameters $a$ and $b$ (see Eq.~\ref{eq:amati2}), one can define the $\chi^2$ as follows:
\begin{equation}
\chi^2 = \sum_{i=1}^{N} \left( \frac{g_{i} -  f(E_{p,i};a,b)}{\sigma_i} \right)^2 \, ,
\label{eq:chisq}
\end{equation}
where $g=\log E_\mathrm{iso}$, is derived from data, and $f$ is the theoretical value of $\log E_\mathrm{iso}$ which can be computed using Eq.~\ref{eq:amati2}. The Amati parameters $a$ and $b$ can be calculated by minimising the $\chi^2$ in Eq.~\ref{eq:chisq}. The following analysis can be used to calculate $\sigma_i$ of $\log E_\mathrm{iso}$. The error in $\log E_\mathrm{iso}$ due to observational uncertainty in $E_\mathrm{iso}$ is $$ dg = dE_{\mathrm{iso}}/E_{\mathrm{iso}} \, . $$ 

But uncertainty in $\log E_\mathrm{iso}$ depends on uncertainty in $\log E_\mathrm{p,i}$, (see Eq.~\ref{eq:amati2}).
$$ df=  b (dE_{\mathrm{p,i}}/E_{\mathrm{p,i}}) \, . $$

Thus, 
\begin{eqnarray}
          \sigma_i^2   = (dg)^2+(df)^2  
          = (dE_{\mathrm{iso}}/E_{\mathrm{iso}})^2 + (b. (dE_{\mathrm{p,i}}/E_{\mathrm{p,i}}))^2 \, . 
\end{eqnarray}

The likelihood, $P(D|H)$, which is the probability of collecting data if the model $H$ is correct, may be expressed in terms of $\chi^2$ as follows:
\begin{equation}
P(D|H(a,b)) \propto \exp{(-\chi^2/2)} \,  ,  
\label{eq:likeli}
\end{equation}
where $\chi^2$ is defined by Eq.~\ref{eq:chisq}. The parameters $a$ and $b$ can be estimated by maximizing likelihood with respect to these parameters. We employ the Bayesian methodology because the above-mentioned method does not provide a direct probability estimate for the Amati parameters. With the help of Bayes' theorem, the posterior probability of the Amati parameters can be calculated as:
\begin{equation}
     P(H(a,b)|D) \propto P(D|H(a,b))\times P(H(a,b)) \, .
    \label{eq:bayes}
\end{equation}
The model's prior probability, $P(H)$, reflects our current understanding of the model. Because priors are subjective, careful consideration should be given to their selection. Another benefit of the Bayesian method is the marginalisation of the nuisance parameters. For instance, marginalization over parameter $a$ leads to the following equation:
\begin{equation}
    P({b}/ \mu) = \int {P(\mu/ a,b)P(a,b)da} \, .
    \label{eq:margin}
\end{equation}
A similar procedure can be used to marginalise over the other parameter, $b$ as needed. To check if the Amati parameters evolve with the redshift, we fit the Amati relation in Eq~\ref{eq:amati2} for each subset using the methods discussed above.
\begin{table}
\begin{center}
\caption{Best-fit values of the Amati parameters, $a$ and $b$ along with $1\sigma$ confidence level. The Bayesian approach has been applied to the complete set of $162$ long GRBs.
\label{tbl:results-bays-full}} \bigskip 
\begin{tabular}{cccccc}
\hline
 $a$ & $ b $ & $\sigma_a$ & $\sigma_b$  \\
\hline
-4.22 & 2.16 & $\pm 0.12$ & $\pm 0.04$\\
\hline\hline
\end{tabular}
\end{center}
\end{table}
\begin{table}
\begin{center}
\caption{Subgroups of GRBs.
\label{tbl:subgroups-1}} \bigskip 
\begin{tabular}{ccccc}
\hline
Group & Number of GRBs & Range of $z$ & Median of $z$   \\
\hline
$\rm{G_1}$ & 71 & 0.033 - 1.489 & 0.86 \\
$\rm{G_2}$ & 91 & 1.52 - 9.3 & 2.48 \\
\hline\hline
\end{tabular}
\end{center}
\end{table}
\begin{table}
\begin{center}
\caption{Best-fit values of the Amati parameters, $a$ and $b$, along with the $1 \sigma$ errors for the subgroups $G_1$ and $G_2$. The Bayesian approach has been used for the estimation.}
\label{tbl:results3} \bigskip 
\begin{tabular}{ccccc}
\hline
Group & $a$ & $b$ \\
\hline
$\rm{G_1}$ & $-3.79\pm 0.145$ & $1.97 \pm 0.06$ \\
$\rm{G_2}$ & $-4.54\pm 0.240$ & $2.26\pm 0.08$ \\
\hline\hline
\end{tabular}
\end{center}
\end{table}

\subsection{Selection Biases}
\label{sec:selection}

The selection effects in the prompt and afterglow observations may arise due to the sensitivity of the instrument as well as the phenomena impacting the detection probability. Due to the instrument's sensitivity and the events that impact the detection likelihood, selection effects in the prompt and afterglow may develop. Some examples of instrumental biases \cite{cow13} are the source localization from the telescope, and the source's position in the sky, especially if it is close to the Sun. Sometimes instrumental biases could be time-dependent, for instance, the learning curve of the instrument can affect the redshift distribution over time. The Malmquist bias arises due to the greater sensitivity of the instruments towards the brighter end of the luminosity function (LF). The knowledge of the LF is necessary to account for this bias.

Another important bias is the redshift desert which is related to the redshift measurement in the range $1.3<z<3$ \cite{cow13}. At a redshift of $z>1$ the optical wavelength in the afterglow shifts to infrared. The sensitivity of CCD drops in this region, and since signal to noise ratio is already poor for faint sources, it becomes difficult to measure the spectra.

The selection biases in the GRB data can lead to false positive results in our investigation. Hence, we perform simulations to test whether the selection biases impact our results. We first simulate a sample of $z$ and $E_{iso}$ following the distribution of the original data. Now, we produce $E_{p,i}$ by fitting the Amati relation to the above simulated sample. We put a lower limit $E_{iso,lim}$ on isotropic-equivalent energy in the simulation, keeping in mind the flux detection limit of the instruments. We divide the sample into low and high-$z$ groups, as was done with the original data. Again, we fit the Amati relation to both groups. This exercise is repeated several times. Best-fit values of the Amati parameters are plotted, and a comparison between the low-$z$ and high-$z$ is made. One can repeat the exercise without implementing the lower limit $E_{iso,lim}$ on isotropic-equivalent energy. A comparison between the above two results infers whether the selection biases impact the earlier investigation result.
\section{Results and Discussion}
\label{sec:results}
We first calculate the $\chi^2$ and minimize it through the Amati parameters a and b. Minimum $\chi^2$ per degree of freedom (dof) is found to be $\chi^2_r=6.32$.
The large value of $\chi^2$ per dof indicates that the errors in the data may have been underestimated. To obtain the direct probability of the parameters, we use the Bayesian approach through Eq.~\ref{eq:bayes}. The best-fit values of the Amati parameters are presented in Table~\ref{tbl:results-bays-full} along with their error bars. Figures~\ref{fig:a-full-post} and \ref{fig:b-full-post} display the posterior probability of each parameter. Bayesian marginalization on parameter $b$ has been applied while calculating the posterior for $a$ and vice-versa. For both parameters, uniform priors in a reasonable range have been considered.

The data is now divided into low-$z$ and high-$z$ groups, $G_1$ and $G_2$, which contain $71$ and $91$ GRBs, respectively. Table~\ref{tbl:subgroups-1} displays the details of these groups. As done previously, we obtain the posterior probability of $a$ and $b$ for each group. Table~\ref{tbl:results3} shows the best-fit values, while figures~\ref{fig:a-grup-post} and \ref{fig:b-grup-post} display the posterior probability distribution for each parameter. The distributions for the parameter, $a$ for the two groups, $G_1$ and $G_2$ are different at around $1.9 \sigma$ level, as seen in Figure~\ref{fig:a-grup-post}. The probability distributions for parameter, $b$ determined for the two groups are also different at more than $2 \sigma$ level, as demonstrated in Figure~\ref{fig:b-grup-post}. 

\subsection{Analysis of Selection effect}
\label{sec:selecion effect}
To confirm whether the differences in low and high redshift parameters are intrinsic, we need to analyse the selection effects which could be involved in the measurement process. Every detector has a lower detection limit of flux, hence the impact is that distant objects detected by the telescope would be intrinsically brighter. To investigate this impact, we use the following scheme, which is a modified version of \cite{Li2007}. 

First, we explore the distribution of the redshift, $z$ and isotropic-equivalent energy, $E_\mathrm{iso}$ of the observed GRBs. They follow a log-normal distribution as shown in Figure~\ref{fig:lognormal}. The mean and the standard deviation for log $z$ are 0.18 and 0.34, respectively. The fit can be expressed as:
\begin{equation}
   	f_1(\mathrm{\log z}) = \frac{1}{\sqrt{2\pi}\sigma_1} \exp\left[-\frac{(\log z
		-\mu_1)^2}{2\sigma_1^2}\right] \;,
	\label{f_logz}
\end{equation}
Similarly, the mean and the standard deviation for log$E_\mathrm{iso}$ are 1.01 and 0.84, respectively, which can be expressed as:
\begin{equation}
   	f_1(\rm{\log E_{iso}}) = \frac{1}{\sqrt{2\pi}\sigma_2} \exp\left[-\frac{(\log E_{iso}
		-\mu_2)^2}{2\sigma_2^2}\right] \;,
	\label{f_logeiso}
\end{equation}
Following \cite{Li2007}, we now simulate the GRB data to understand the selection effects. We generate $1000$ sets of GRB samples, each containing $1000$ values of $z$ and $E_{iso}$ using the log-normal distribution of Eq.~\ref{f_logz} and Eq.~\ref{f_logeiso}, respectively. The distribution's mean and standard deviation are the same as those of the observed GRBs in Fig.~\ref{fig:lognormal}. Keeping in mind that every detector has a lower flux detection limit, we put a lower limit of $E_\mathrm{iso}$ in the simulated data. 
In figure~\ref{fig:fbol}, we plot the isotropic-equivalent energy of the 162 GRBs in our sample. The plot clearly shows that the (observed minimum) isotropic energy is correlated with the redshift. The solid line in the plot shows the limit corresponding to a typical value for the bolometric fluence, $F_{\mathrm bol,lim} = {10^{-9} {\mathrm erg cm}^{-2}}$, which fairly represents the lower limit of the isotropic-equivalent energy of a detectable burst at redshift $z$ in our sample. At the redshift $z$, the lower limit on $E_{\mathrm{iso}}$ can be given as: 
\begin{equation}
   	E_{\mathrm{iso,lim}} = 4\pi {D_{\mathrm{com}}}^2(1+z)F_{\mathrm{bo,lim}}
	\label{eq:limit}
\end{equation}
where ${D_\mathrm{com}}$ is the comoving distance to the burst. Having the set of $($z$, {E_\mathrm{iso}})$ pairs, we generate ${E_\mathrm{p,i}}$ for each pair as:
\begin{equation}
 	f_2(y) = {\frac{1}{\sqrt{2\pi}\sigma_3}} exp\left[-\frac{(y-sx-p)^2}{2{\sigma_3}^2}\right] \; ,
	\label{f_y}
\end{equation}
where $x=log(E_\mathrm{iso})$ and $y=log(E_\mathrm{p,i})$. Here, $s$ and $p$ represent the slope and the intercept for the relation, $y=sx+p$. Having the sample of $($z$,{E_\mathrm{iso}},{E_\mathrm{p,i}})$, we divide them into low and high-$z$ groups with the boundary as $z=1.5$. Now, we determine the Amati parameters for both groups in the sample. This process is repeated one thousand times, and the distribution of the Amati parameters for these samples is shown in  Figure~\ref{fig:withselectioneffect}. The graph shows that most of the best-fit values of the two groups overlap. In fact, the average value of the Amati parameters for the low redshift group is (-4.59, 2.21) and for the high redshift group is (-4.56, 2.20), which do not differ much from each other. 
This conflicts with figures~\ref{fig:a-grup-post} and \ref{fig:b-grup-post} indicating that the $2\sigma$ difference in Amati parameters for low and high-$z$ samples is not expected. To explore the matter further, we turn off the lower limit on $E_{iso}$, defined in Eq.~\ref{eq:limit} and again simulate one thousand samples of $z$ and $E_{iso}$, each containing $1000$ data points. Similar to the preceding instance, Eq.~\ref{f_y} is used to generate $E_{p,i}$ corresponding to each pair of $z$ and $E_{iso}$. Each sample is again divided into low and high-$z$ groups, and best-fit is obtained. The best-fit values have been plotted in Figure~\ref{fig:withoutselectioneffect}. Similar to Figure~\ref{fig:withselectioneffect}, most of the values overlap, and there is hardly any difference in the average values of the Amati parameters for low and high-$z$ groups. 
\section{Conclusion}
\label{sec:conclusion}
The goal of the present work was to study the possible redshift evolution of the Amati relation, which is a correlation between the isotropic-equivalent energy, $E_\mathrm{iso}$ and the peak energy, $E_\mathrm{p,i}$. If the GRB properties evolve with redshift, it will be reflected in the Amati parameters. Alternatively, the selection biases could also lead to false detection of the evolution. We have applied the Bayesian approach to estimate the Amati parameters $a$ and $b$. The data set containing 162 long GRBs up to a redshift of $9.3$ \cite{Demianski17} was divided into two groups below and above redshift $z=1.5$. The selection of $z\sim 1.5$ is appropriate as the quenching activity in galaxy clusters and a change in star formation rate has been indicated around this redshift \cite{Nantais2016,Krumholz2012,Zhiyuan2018,Rychard2020}. Our analysis in \S~\ref{sec:results} shows that the best-fit values of Amati parameters for the two groups of GRBs are quite different. The values don't match at $\sim 2 \sigma$ level. In order to determine whether the disparity may be caused by the presence of selection biases, we simulate the GRB data. Our analysis ruled out the selection bias as the possible cause of the mismatch. Thus, it seems reasonable to infer that the mismatch in the low and high-$z$ values of the Amati parameters could be linked with the evolution of GRB with redshift. It is in line with recent studies \cite{Lloyd2019,Maria2021} which suggest a strong evolution of parameters like $E_{iso}$, $T_{90}$ and GRB luminosity. Understanding the possible evolution might be beneficial in the calibration of GRBs as secondary distance indicators. This completely alters the high redshift cosmology.
\section*{acknowledgments}
Meghendra Singh thanks DMRC for Support Darshan Singh thanks the compeers of GD Goenka University for eternal assistance. 
\bibliographystyle{mnras}
\bibliography{main} 
\end{document}